\documentclass[12pt]{iopart}
\usepackage{epsfig}  
\begin{document}

\title[The LATOR mission]{The Laser Astrometric Test of Relativity (LATOR) Mission}

\author{Slava G. Turyshev\dag, Michael Shao\dag, and Kenneth Nordtvedt\ddag\  
\footnote[3]{To
whom correspondence should be addressed (turyshev@jpl.nasa.gov)}
}

\address{\dag\ Jet Propulsion Laboratory, California Institute of  Technology, 4800 Oak Grove Drive, Pasadena, CA 91109 USA}

\address{\ddag\ Northwest Analysis, 118 Sourdough Ridge Road, Bozeman, MT 59715 USA}

\begin{abstract}
This paper discusses the motivation and general design elements of a new fundamental physics experiment that will test relativistic gravity at the accuracy better than the effects of the second order in the gravitational field strength, $\propto G^2$. The Laser Astrometric Test Of Relativity (LATOR) mission uses laser interferometry between two micro-spacecraft whose lines of sight pass close by the Sun to accurately measure deflection of light in the solar gravity.  The key element of the experimental design is a redundant geometry optical truss provided by a long-baseline (100 m) multi-channel stellar optical interferometer placed on the International Space Station (ISS). The spatial interferometer is used for measuring the angles between the two spacecraft and for orbit determination purposes. In Euclidean geometry, determination of a triangle's three sides determines any angle therein; with gravity changing the optical lengths of sides passing close by the Sun and deflecting the light, the Euclidean relationships are overthrown. The geometric redundancy enables LATOR to measure the departure from Euclidean geometry caused by the solar gravity field to a very high accuracy.

LATOR will not only improve the value of the parameterized post-Newtonian (PPN) parameter $\gamma$ to unprecedented levels of accuracy of 1 part in 10$^{8}$, it will also reach ability to measure effects of the next post-Newtonian order ($\propto G^2$) of light deflection resulting from gravity's intrinsic non-linearity.  The solar quadrupole moment parameter, $J_2$, will be measured with high precision, as well as a variety of other relativistic effects including Lense-Thirring precession.  LATOR will lead to very robust advances in the tests of Fundamental physics: this mission could discover a violation or extension of general relativity, or reveal the presence of an additional long range interaction in the physical law.  There are no analogs to the LATOR experiment; it is unique and is a natural culmination of solar system gravity experiments.

\end{abstract}



\maketitle

\section{Introduction}

Einstein's general theory of relativity (GR) began with its empirical success in 1915 by explaining the anomalous perihelion precession of Mercury's orbit, using no adjustable theoretical parameters.  Shortly thereafter, Eddington's 1919 observations of star lines-of-sight during a solar eclipse confirmed the doubling of the deflection angles predicted by GR as compared to Newtonian-like and Equivalence Principle arguments.  From these beginnings, GR has been verified at ever higher accuracy. Thus, microwave ranging to the Viking Lander on Mars yielded accuracy  $\sim$0.1\% in the tests of GR \cite{viking_shapiro1,viking_reasen,viking_shapiro2}. The astrometric observations of quasars on the solar background performed with Very-Long Baseline Interferometry (VLBI) improved the accuracy of the tests of gravity to $\sim$0.03\% \cite{eubanks97,RoberstonCarter91,Lebach95}. 
Lunar Laser Ranging (LLR),  a continuing legacy of the Apollo program, provided $\sim$0.01\% verification of GR via precision measurements of the lunar orbit \cite{Ken_LLR68,Ken_LLR91,Ken_LLR98,Ken_LLR30years99,Ken_LLRgravitomagnetic88,Ken_LLR_PPNprobe03,JimSkipJean96,Williams_etal_2001}. Finally, the recent experiments with the Cassini spacecraft improved the accuracy of the tests to $\sim$0.0023\% \cite{iess_etal_1999,cassini_ber, cassini_and}. As a result, by now not only the ``non-relativistic,'' Newtonian regime of GR is well understood, but also the first ``post-Newtonian'' approximation is also well-studied, making GR the standard theory of gravity when astrometry and spacecraft navigation are concerned. 

However, the continued inability to merge gravity with quantum mechanics, and recent cosmological observations indicate that the pure tensor gravity of GR needs modification.  Progress in scalar-tensor extensions of gravity which are consistent with present cosmological models \cite{damour_nordtvedt1,damour_nordtvedt2,DPV02,DPV02b} motivate new searches for very small deviations of relativistic gravity in the solar system, at levels of 10$^{-5}$ to 10$^{-7}$ of the post-Newtonian effects or essentially to achieve accuracy that is compatible to the size of the effects of the second order in the gravitational field strength ($\propto G^2$).  This will require a several order of magnitude improvement in experimental precision from present tests. The ability to measure the first order light deflection term at the accuracy comparable with the effects of the second order is of the utmost importance for the gravitational theory and is the challenge for the 21st century fundamental physics. 

\begin{table}[t!]
\caption{Comparable sizes of various light deflection effects in the solar gravity field. The value of deflection angle calculated on the limb of the Sun ($b=R_\odot$); the corresponding delay is given for a 100 m interferometric baseline proposed for LATOR. \label{tab:eff}}
\vskip 5pt
\begin{tabular}{|c|c|c|c|} \hline  
    & & Deflection &\\[-3pt]
    Effect  & 
     Analytical Form      & 
      angle, $\mu$as  &
     Delay, pm \\ 
\hline \hline
&&&\\[-12pt]
   First  Order  &
   $2(1+\gamma )\frac{M}{b}$ & 
   $1.75\times 10^6$   & 
   $0.849 $~mm     \\  
&&&\\[-12pt] \hline 
&&&\\[-12pt]
   Second Order   &
   $\{[2(1+  \gamma)-\beta+\frac{3}{4} \delta]\pi- 
2(1+\gamma)^2\}\frac{M^2}{b^2} $
   &  3.5   
   & 1697      \\ 
&&&\\[-12pt] \hline 
&&&\\[-12pt]
   Frame-Dragging   &
   $\pm 2(1+\gamma)\frac{J}{b^2}$ & 
   $\  \pm0.7$   &     $\pm339$ \\ 
&&&\\[-12pt] \hline 
&&&\\[-12pt]
   Solar Quadrupole  &
   $2(1+\gamma )J_2\frac{M}{b^3}$ & 
     0.2   &   97  \\[4pt] 
\hline 
\end{tabular} 
\end{table}

When the light deflection in solar gravity is concerned, the magnitude of the first order effect as predicted by GR for the light ray just grazing the limb of the Sun is $\sim1.75$ arcsecond (asec) (for more details see Table \ref{tab:eff}). (Note that 1 arcsec $\simeq5~\mu$rad; when convenient, below we will use the units of radians and arcseconds interchangeably.) The effect varies inversely with the impact parameter. The second order term is almost six orders of magnitude smaller resulting in  $\sim 3.5$ microarcseconds ($\mu$as) light deflection effect, and which falls off inversely as the square of the light ray's impact parameter \cite{EpsteinShapiro80,FishbachFreeman80,RichterMatzner82_1,RichterMatzner82_2,Ken_2PPN_87}. The relativistic frame-dragging term\footnote{Gravitomagnetic frame dragging is the effect in which both the orientation and trajectory of objects in orbit around a body are altered by the gravity of the body's rotation.  It was studied by Lense and Thirring in 1918.} is $\pm 0.7 ~\mu$as, and contribution of the solar quadrupole moment, $J_2$, is sized as 0.2 $\mu$as (using theoretical value of the solar quadrupole moment $J_2\simeq10^{-7}$ ). The small magnitudes of the effects emphasize the fact that, among the four forces of nature, gravitation is the weakest interaction; it acts at very long distances and controls the large-scale structure of the universe, thus, making the precision tests of gravity a very challenging task. 

The LATOR  mission will directly address the challenges discussed above. The test will be performed in the solar gravity field using optical interferometry between two micro-spacecraft.  Precise measurements of the angular position of the spacecraft will be made using a fiber coupled multi-chanelled optical interferometer on the ISS with a 100 m baseline. The primary objective of LATOR will be to measure the gravitational deflection of light by the solar gravity to accuracy of 0.1 picoradians (prad) ($\sim0.02 ~\mu$as), which corresponds to $\sim$10 picometers (pm) on a 100 m interferometric baseline. 

In conjunction with laser ranging among the spacecraft and the ISS, LATOR will allow measurements of the gravitational deflection by a factor of more than 3,000 better than had recently been accomplished with the Cassini spacecraft. In particular, this mission will not only measure the key PPN parameter $\gamma$ to unprecedented levels of accuracy of one part in 10$^8$, it will also reach ability to measure the next post-Newtonian order ($\propto G^2$) of light deflection resulting from gravity's intrinsic non-linearity. As a result, this experiment will measure values of other PPN parameters such as parameter $\delta$ to 2 part in $10^3$ (never measured before; see Eq.~(\ref{eq:metric}) and discussion thereafter), the solar quadrupole moment parameter $J_2$ to 1 part in 20, and the frame dragging effects on light due to the solar angular momentum to precision of 1 parts in $10^2$.

The LATOR mission technologically is a very sound concept; all technologies that are needed for its success have been already demonstrated as a part of the JPL's Space Interferometry Mission (SIM) development. (In fact, accuracy of 5 pm was already demonstrated in our SIM-related studies.)  The LATOR concept arose from several developments at NASA and JPL that initially enabled optical astrometry and metrology, and also led to developing expertize needed for the precision gravity experiments. Technology that has become available in the last several years such as low cost microspacecraft, medium power highly efficient solid state and fiber lasers, and the development of long range interferometric techniques make possible an unprecedented factor of 3,000 improvement in this test of GR possible. This mission is unique and is the natural next step in solar system gravity experiments which fully exploits modern technologies.

LATOR will lead to very robust advances in the tests of fundamental physics:  this mission could discover a violation or extension of GR, or reveal the presence of an additional long range interaction in the physical law.  With this mission testing theory to several orders of magnitude higher precision, finding a violation of GR or discovering a new long range interaction could be one of this era's primary steps forward in fundamental physics. There are no analogs to the LATOR experiment; it is unique and is a natural culmination of solar system gravity experiments.  

This paper organized as follows: Section \ref{sec:sci_mot} discusses the theoretical framework and science motivation for the precision gravity tests that recently became available.  Section \ref{sec:lator_description} provides an overview for the LATOR experiment including the current mission design.  Section \ref{sec:error_bud} discusses a preliminary  error budget. In Section \ref{sec:conc} we discuss the next steps that will taken in the development of the LATOR mission.

\section{Scientific Motivation}
\label{sec:sci_mot}
\subsection{PPN Parameters and Their Current Limits}

Generalizing on a phenomenological parameterization of the gravitational metric tensor field which Eddington originally developed for a special case, a method called the parameterized post-Newtonian (PPN) metric has been developed (see \cite{Ken_LLR68,Ken_LLR91,Ken_2PPN_87,Ken_EqPrinciple68,Will_book93,WillNordtvedt72}).
This method  represents the gravity tensor's potentials for slowly moving bodies and weak interbody gravity, and it is valid for a broad class of metric theories including GR as a unique case.  The several parameters in the PPN metric expansion vary from theory to theory, and they are individually associated with various symmetries and invariance properties of underlying theory.  Gravity experiments can be analyzed in terms of the PPN metric, and an ensemble of experiments will determine the unique value for these parameters, and hence the metric field, itself.

In locally Lorentz-invariant theories the expansion of the metric field for a single, slowly-rotating gravitational source in PPN parameters is given by:
{}
\begin{eqnarray}
\label{eq:metric}
g_{00}&=&1-2\frac{GM}{c^2r}\Big(1-J_2\frac{R^2}{r^2}\frac{3\cos^2\theta-1}{2}\Big)+2\beta\Big(\frac{GM}{c^2r}\Big)^2+{\cal O}(c^{-5}),\nonumber\\
g_{0i}&=& 2(\gamma+1)\frac{G[\vec{J}\times \vec{r}]_i}{c^3r^3}+
{\cal O}(c^{-5}),\\\nonumber
g_{ij}&=&-\delta_{ij}\Big[1+2\gamma \frac{GM}{c^2r} \Big(1-J_2\frac{R^2}{r^2}\frac{3\cos^2\theta-1}{2}\Big)+\frac{3}{2}\delta \Big(\frac{GM}{c^2r}\Big)^2\Big]+{\cal O}(c^{-5}),   
\end{eqnarray}

\noindent where $M$ and $\vec J$ being the mass and angular momentum of the Sun, $J_2$ being the quadrupole moment of the Sun and $R$ being its radius.  $r$ is the distance between the observer and the center of the Sun.  $\beta, \gamma, \delta$ are the PPN parameters and in GR they are all equal to $1$. The $M/r$ term in the $g_{00}$ equation is the Newtonian limit; the terms multiplied by the post-Newtonian parameters $\beta, \gamma$,  are post-Newtonian terms. The term multiplied by the post-post-Newtonian parameter  $\delta$ also enters the calculation of the relativistic light deflection \cite{Ken_cqg96}.

This PPN expansion serves as a useful framework to test relativistic gravitation in the context of the LATOR mission. In the special case, when only two PPN parameters ($\gamma$, $\beta$) are considered, these parameters have clear physical meaning. Parameter $\gamma$  represents the measure of the curvature of the space-time created by a unit rest mass; parameter  $\beta$ is a measure of the non-linearity of the law of superposition of the gravitational fields in the theory of gravity. GR, which corresponds to  $\gamma = \beta$  = 1, is thus embedded in a two-dimensional space of theories. The Brans-Dicke is the best known theory among the alternative theories of gravity.  It contains, besides the metric tensor, a scalar field and an arbitrary coupling constant $\omega$, which yields the two PPN parameter values $\gamma = (1+ \omega)/(2+ \omega)$, and $\beta$  = 1.  More general scalar tensor theories yield values of $\beta$ different from one.

The PPN formalism has proved to be a versatile method to plan gravitational experiments in the solar system and to analyze the data obtained 
\cite{Ken_LLR68,Ken_LLR91,Ken_2PPN_87,Ken_EqPrinciple68,Will_book93,WillNordtvedt72,Ken_cqg96,ken_icarus95,Will_reviews,anderson_mars96,Bender_LLR97}.
Different experiments test different combinations of these parameters (for more details, see \cite{Will_book93}). The most precise value for the PPN parameter  $\gamma$ is at present given by the Cassini mission \cite{cassini_ber} as: $\gamma -1 = (2.1\pm2.3)\times10^{-5}$. The secular trend of Mercury's perihelion, when described in the PPN formalism, depends on another linear combination of the PPN parameters $\gamma$  and  $\beta$ and the quadrupole coefficient $J_{2}$ of the solar gravity field:  $\lambda_\odot = (2 + 2\gamma -\beta)/3 + 0.296\times J_{2}\times 10^4$. The combination of parameters $\lambda_\odot$, was obtained with Mercury ranging data as $\lambda_\odot = 0.9996\pm 0.0006$ \cite{Pitjeva93}. The PPN formalism has also provided a useful framework for testing the violation of the Strong Equivalence Principle (SEP) for gravitationally bound bodies.  In that formalism, the ratio of passive gravitational mass $M_G$ to inertial mass $M_I$ of the same body is given by $M_G/M_I = 1 -\eta U_G/(M_0c^2)$, where $M_0$ is the rest mass of this body and $U_G$ is the gravitational self-energy. The SEP violation is quantified by the parameter $\eta$, which is expressed in terms of the basic set of PPN parameters by the relation $\eta = 4\beta-\gamma-3$. 
Analysis of planetary ranging data recently yielded an independent determination of parameter $\gamma$ \cite{Williams_etal_2001,Anderson_etal_GRtests02,AndersonWilliams01}: $|\gamma -1| = 0.0015 \pm 0.0021$; it also gave   with accuracy at the level of $|\beta -1| = -0.0010 \pm 0.0012$. 
Finally, with LLR finding that Earth and Moon fall toward the Sun at rates equal to 1.5 parts in 10$^{13}$, even in a conservative scenario where a composition dependence of acceleration rates masks a gravitational self energy dependence $\eta$ is constrained to be less than 0.0008 \cite{AndersonWilliams01}; without such accidental cancelation the $\eta$ constraint improves to 0.0003. Using the recent Cassini result \cite{cassini_ber} on the PPN  $\gamma$, the parameter $\beta$ was measured as $\beta-1=(0.9\pm1.1)\times 10^{-4}$ from LLR. The next order PPN parameter $\delta$ has not yet been measured though its value can be inferred from other measurements.

Therefore, a careful  design of a designated gravitational experiment may offer a significant accuracy improvement in the relativistic gravity tests in the weak-field limit of the solar system. 
In fact, the technology has advanced to the point that one can consider carrying out direct tests in a weak field to second order in the field strength parameter ($\propto G^2$). Although any measured anomalies in first or second order metric gravity potentials will not determine strong field gravity, they would signal that modifications in the strong field domain will exist.  The converse is perhaps more interesting:  if to high precision no anomalies are found in the lowest order metric potentials, and this is reinforced by finding no anomalies at the next order, then it follows that any anomalies in the strong gravity environment are correspondingly quenched under all but exceptional circumstances (for example, a mechanism of a
``spontaneous-scalarization'' that, under certain circumstances, may exist in tensor-scalar theories \cite{Damour_EFarese93}). 

We shall now discuss the recent motivations for the precision gravity experiments.

\subsection{Motivations for Precision Gravity Experiments}
\label{sec:mot}

Recently considerable interest has been shown in the physical processes occurring in the strong gravitational field regime. It should be noted that GR and some other alternative gravitational theories are in good agreement with the experimental data collected from the relativistic celestial mechanical extremes provided by the relativistic motions in the binary millisecond pulsars.  However, many modern theoretical models, which include GR as a standard gravity theory, are faced with the problem of the unavoidable appearance of space-time singularities. It is generally suspected that the classical description, provided by GR, breaks down in a domain where the curvature is large, and, hence, a proper understanding of such regions requires new physics. 

The continued inability to merge gravity with quantum mechanics suggests that the pure tensor gravity of GR needs modification or augmentation. The tensor-scalar theories of gravity, where the usual general relativity tensor field coexists with one or several long-range scalar fields, are believed to be the most promising extension of the theoretical foundation of modern gravitational theory. The superstring, many-dimensional Kaluza-Klein, and inflationary cosmology theories have revived interest in the so-called `dilaton fields', i.e. neutral scalar fields whose background values determine the strength of the coupling constants in the effective four-dimensional theory. The importance of such theories is that they provide a possible route to the quantization of gravity and unification of physical law. Although the scalar fields naturally appear in the theory, their inclusion predicts different relativistic corrections to Newtonian motions in gravitating systems. These deviations from GR lead to a violation of the Equivalence Principle (either weak or strong or both), modification of large-scale gravitational phenomena, and generally lead to space and time variation of physical ``constants.'' As a result, this progress has provided new strong motivation for high precision relativistic gravity tests.

\begin{figure}[t!]
\begin{center}
\vskip 5pt
\psfig{figure=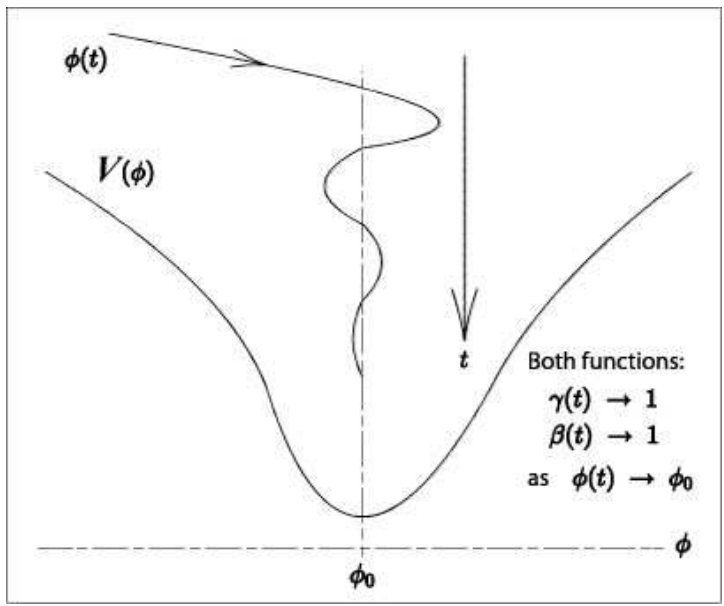,width=95mm}
\caption{Typical cosmological dynamics of a background scalar field is shown in the case when that field's coupling function to matter, $V(\phi)$, has an attracting point $\phi_0$. The strength of the scalar interaction's coupling to matter is proportional to the derivative (slope) of the coupling function, so it weakens as the attracting point is approached, and both the Eddington parameters $\gamma$ and $\beta$ (and all higher structure parameters as well)  approach their pure tensor gravity values in this limit \cite{Damour_EFarese96,damour_nordtvedt1,damour_nordtvedt2,DPV02,DPV02b}.  But a small residual scalar gravity should remain today because this dynamical process is not complete, and that is what experiment seeks to find.
 \label{fig:attract}}
\end{center}
\end{figure} 


The recent theoretical findings suggest that the present agreement between Einstein's theory and experiment might be naturally compatible with the existence of a scalar contribution to gravity. In particular, Damour and Nordtvedt \cite{damour_nordtvedt1,damour_nordtvedt2} (see also 
\cite{DamourGG90,DamourTaylor92,DamourPolyakov94_1,DamourPolyakov94_2} for non-metric versions of this mechanism and \cite{DPV02,DPV02b} for the recent summary of a dilaton-runaway scenario) have found that a scalar-tensor theory of gravity may contain a `built-in' cosmological attractor mechanism towards GR.  A possible scenario for cosmological evolution of the scalar field was given in \cite{Ken_LLR_PPNprobe03,damour_nordtvedt2}. Their speculation assumes that the parameter  $\frac{1}{2}(1-\gamma)$  was of order of 1 in the early universe, at the time of inflation, and has evolved to be close to, but not exactly equal to, zero at the present time (Figure \ref{fig:attract} illustrates this mechanism in more details). The expected deviation from zero may be of order of the inverse of the redshift of the time of inflation, or somewhere between 1 part per $10^5$ and 1 part per $10^7$ depending on the total mass density of the universe:  $1-\gamma \sim 7.3 \times 10^{-7}(H_0/\Omega_0^3)^{1/2}$, where $\Omega_0$ is the ratio of the current density to the closure density and $H_0$ is the Hubble constant in units of 100 km/sec/Mpc. Compared to the cosmological constant, these scalar field models are consistent with the supernovae observations for a lower matter density, $\Omega_0\sim 0.2$, and a higher age, $(H_0 t_0) \approx 1$. If this is indeed the case, the level $1-\gamma \sim 10^{-6}-10^{-7}$ would be the lower bound for the present value of PPN parameter $\gamma$ \cite{damour_nordtvedt1,damour_nordtvedt2}. 

More recently, Damour et al. have given a
new estimation of $|1-\gamma|$, within the framework compatible with
string theory and modern cosmology, which basically confirms
the previous result \cite{DPV02,DPV02b}. This recent analysis discusses a scenario when a composition-independent coupling of dilaton to hadronic matter produces detectable deviations from GR in experiments conducted with accuracy better than $2\times10^{-5}$. This work assumes only some general property of the coupling functions (for large values of the field, i.e. for an ``attractor at infinity'') and then only assume that $(1-\gamma)$ is of order of one at the beginning of the controllably classical part of inflation.
It was shown  in \cite{DPV02b} that one can relate the present value of $(1-\gamma)$ to the cosmological density fluctuations.
For the simplest inflationary potentials (favored by WMAP mission, i.e. $m^2 \chi^2$ \cite{wmap,[4c]}) Damour et al. found that the present value of $(1-\gamma)$ could be just below $10^{-7}$.
In particular, within this framework $|1-\gamma|\simeq-2\alpha^2_{had}$, where $\alpha_{had}$ is the dilaton coupling to hadronic matter. Its value depends on the model taken for the inflation potential $V(\chi)\propto\chi^n$, with $\chi$ being the inflation field; the level of the expected deviations from GR is $\sim0.5\times10^{-7}$  for $n = 2$ \cite{DPV02b}. Note that these predictions are based on the work in scalar-tensor extensions of gravity which are consistent with, and indeed often part of, present cosmological models. The analyses discussed above not only motivate new searches for very small deviations of relativistic gravity in the solar system, they also predict that such deviations are currently present in the range from 10$^{-5}$ to $\sim 5\times 10^{-8}$ of the post-Newtonian effects and, thus, should be easily detectable with LATOR.

There is now multiple evidence indicating that 70\% of the critical density of the universe is in the form of a ``negative-pressure'' dark energy component; there is no understanding as to its origin and nature. The fact that the expansion of the universe is currently undergoing a period of acceleration now seems rather well tested: it is directly measured from the light-curves of several hundred type Ia supernovae \cite{perlmutter99,Riess_supernovae98,[3c]}, and independently inferred from observations of CMB by the WMAP satellite \cite{[4c]} and other CMB experiments \cite{[6c],[5c]}. Cosmic speed-up can be accommodated within GR by invoking a mysterious cosmic fluid with large negative pressure, dubbed dark energy. The simplest possibility for dark energy is a cosmological constant; unfortunately, the smallest estimates for its value are 55 orders of magnitude too large (for reviews see \cite{Carroll_01,PeeblesRatra03}). Most of the theoretical studies operate in the shadow of the cosmological constant problem, the most embarrassing hierarchy problem in physics. This fact has motivated a host of other possibilities, most of which assume $\Lambda=0$, with the dynamical dark energy being associated with a new scalar field (see \cite{[carroll]} and references therein). However, none of these suggestions is compelling and most have serious drawbacks. Given the challenge of this problem, a number of authors considered the possibility that cosmic acceleration is not due to some kind of stuff, but rather arises from new gravitational physics (see discussion in 
\cite{Carroll_01,PeeblesRatra03,[carroll],Carroll_HT_03}).
In particular, some extensions to GR in a low energy regime \cite{[carroll]} were shown to predict an experimentally consistent universe evolution  without the need for dark energy. These dynamical models are expected to produce measurable contribution to the parameter $\gamma$  in experiments conducted in the solar system also at the level of $1-\gamma \sim 10^{-7}-10^{-9}$, thus further motivating the relativistic gravity research. Therefore, the PPN parameter $\gamma$ may be the only key parameter that holds the answer to most of the questions discussed above. Also an anomalous parameter $\delta$ will most likely be accompanied by a `$\gamma$ mass' of the Sun which differs from the gravitational mass of the Sun and therefore will show up as anomalous $\gamma$ (see discussion in \cite{Ken_LLR_PPNprobe03}).

Finally, even in the solar system, GR still faces challenges. There is the long-standing problem of the size of the solar quadrupole moment and its possible effect on the relativistic perihelion precession of Mercury (see review in \cite{Will_book93}). The interest in lies in the study of the behavior of the solar quadrupole moment versus the radius and the heliographic latitudes. This solar parameter has been very often neglected in the past, because it was rather difficult to determine an accurate value. The improvement of our knowledge of the accuracy of $J_2$ is certainly due to the fact that, today, we are able to take into account the differential rotation with depth. In fact, the quadrupole moment plays an important role in the accurate computation of several astrophysical quantities, such as the ephemeris of the planets or the general relativistic prediction for the precession of the perihelion of Mercury and other minor planets such as Icarus.  Finally, it is necessary to accurately know the value of the quadrupole moment to determinate the shape of the Sun, that is to say its oblateness. Solar oblateness measurements by Dicke and others in the past gave conflicting results for $J_2$ (reviewed on p. 145 of \cite{[16cw]}). A measurement of solar oblateness with the balloon-borne Solar Disk Sextant gave a quadrupole moment on the order of $2\times 10^{-7}$ \cite{lydon}. Helioseismic determinations using solar oscillation data have since implied a small value for $J_2$, on the order of $\sim 10^{-7}$, that is consistent with simple uniform rotation \cite{Will_book93,Brown,GoughToomre_InteriorSun91}. However, there exist uncertainties in the helioseismic determination for depths below roughly 0.4 $R_\odot$ which might permit a rapidly rotating core. LATOR can measure $J_2$ (with accuracy sufficient to put this issue to rest). 

In summary, there are a number of theoretical reasons to question the validity of GR. Despite the success of modern gauge field theories in describing the electromagnetic, weak, and strong interactions, it is still not understood how gravity should be described at the quantum level. In theories that attempt to include gravity, new long-range forces can arise in addition to the Newtonian inverse-square law. Even at the purely classical level, and assuming the validity of the Equivalence Principle, Einstein's theory does not provide the most general way to generate the space-time metric. Regardless of whether the cosmological constant should be included, there are also important reasons to consider additional fields, especially scalar fields.   The LATOR mission is designed to address theses challenges.

\subsection{LATOR vs Other Gravity Experiments}

Prediction of possible deviation of PPN parameters from the general relativistic values provides a robust theoretical paradigm and constructive guidance for experiments that would push beyond the present empirical upper bound on  the PPN parameter $\gamma$  of $\gamma-1=(2.1\pm2.3)\times10^{-5}$ obtained by recent conjunction experiments with Cassini spacecraft \cite{cassini_ber}. {In addition, any experiment pushing the present upper bounds on $\beta$ (i.e. $|\beta-1| < 5\times 10^{-4}$ from \cite{Williams_etal_2001,AndersonWilliams01}) will also be of great interest.} 

The Eddington parameter $\gamma$, whose value in general relativity is unity, is perhaps the most fundamental PPN parameter, in that $(1-\gamma)$ is a measure, for example, of the fractional strength of the scalar gravity interaction in scalar-tensor theories of gravity \cite{Damour_EFarese96}.  Within perturbation theory for such theories, all other PPN parameters to all relativistic orders collapse to their general relativistic values in proportion to $(1-\gamma)$. This is why measurement of the first order light deflection effect at the level of accuracy comparable with the second-order contribution would provide the crucial information separating alternative scalar-tensor theories of gravity from GR \cite{Ken_2PPN_87} and also to probe possible ways for gravity quantization and to test modern theories of cosmological evolution discussed in the previous section. The LATOR mission is designed to directly address this issue with an unprecedented accuracy.

Tests of fundamental gravitational physics feature prominently among NASA and ESA goals, missions, and programs. Among the future missions that will study the nature of gravity, we discuss here the missions most relevant to LATOR science:

\begin{itemize}
\item 
Configuration similar to the geometry of the Cassini conjunction experiments may be utilized for the microwave ranging between the Earth and a lander on Mars. If the lander were to be equipped with a Cassini-class dual X- and Ka-band communication system, the measurement of the PPN parameter $\gamma$ is possible with accuracy of $\sim$1 part in 10$^6$. As oppose to any scenario involving ranging out to the Martian vicinity, the LATOR mission will not be affected by the complexity of the asteroid modeling problem.

\item 
An ambitious test of one of the foundations of GR -- the Equivalence Principle -- is proposed for the STEP (Space Test of Equivalence Principle) mission, that is currently being developed by the Stanford GB-P group. The experiment will test the composition independence of gravitational acceleration for laboratory-sized bodies by searching for a violation of the EP with a fractional acceleration accuracy of $\Delta a/a\sim 10^{-18}$ \cite{step,step2}. STEP will be able to test very precisely for any non-metric, long range interactions in physical law, however the results of this mission will say nothing about the metric component of gravity itself. The LATOR mission is designed specifically to test the metric nature of the gravitational interaction.

\item 
The SORT (Solar Orbit Relativity Test) mission concept proposes to use laser pulses and a drag-free spacecraft aided with a precision clock orbiting around the Sun to precisely measure $\gamma$ and $J_2$ (solar quadrupole moment)  \cite{sort,veillet93,veillet94}. SORT would combine a time-delay experiment (via laser signals sent from the Earth and recorded by precise clocks on board two satellites orbiting the Sun) with a light-deflection experiment (interferometric measurement on Earth of the angle between the two light signals emitted from the satellites) \cite{sort,Re1999}. As such, SORT would attempt to measure parameter $\gamma$ with accuracy of 1 part in 10$^6$. As opposed to the SORT mission, the LATOR experiment will relay on the redundant geometry formed by the three flight segments (two spacecraft behind the sun and the ISS) and will not utilize either ultra-stable clocks nor ground-based interferometry.

\item 
The ESA's  BepiColombo mission will explore the planet Mercury with equipment allowing an extremely accurate tracking.  This mission will conduct relativity experiments including the study of Mercury's perihelion advance and the relativistic light propagation near the Sun. The BepiColombo mission will enable achievement of the following accuracies: $\sigma_\gamma\simeq 2\times10^{-6}$, $\sigma_\beta\simeq 2\times10^{-6}$ and $\sigma_{J_2}\simeq 2\times10^{-9}$ in measuring the main post-Newtonian parameters \cite{BepiColombo2002}. While a very impressive mission design, its expected accuracy is at least two orders of magnitude worse than that expected from LATOR.  The LATOR mission is a designated relativity mission and it is designed to test solar gravity with accuracy better than $10^{-8}$. 

\item 
We stress that the future optical interferometers in space
such as NASA's SIM (Space Interferometry Mission) and ESA's GAIA (Global Astrometric Interferometer for Astrophysics \cite{gaia1995}) would provide improvement in measurement of relativistic parameters  as a  by-product of their astrometric  program. Thus, SIM will be able to reach accuracy of $\sim 10^{-6}$ in measuring PPN parameter $\gamma$.
GAIA may potentially reach the accuracy of $10^{-5}-5\times 10^{-7}$ in measuring the $\gamma$ \cite{gaia2003}.

\item 
A mission concept aiming to reach comparable accuracies in the tests of relativistic gravity in the solar system had been studied in \cite{astrod02} (see also references therein), and \cite{astrod03}. 
The Astrodynamical Space Test of Relativity using Optical Devices (ASTROD) is an ambitious mission concept that utilizes three drag-free spacecraft — one near L1/L2 point, one with an inner solar orbit and one with an outer solar orbit, ranging coherently with one another using lasers to test relativistic gravity, to measure the solar system and to detect gravitational waves.
The mission may improve the accuracy of determination of the PPN parameter $\gamma$ to $\sim 10^{-7}$ for mini-ASTROD and to $\sim 5\times 10^{-9}$ for a full-scale version \cite{astrod02}. Because of the technological complexity, the launch of an ASTROD-like mission is expected after 2020.
\end{itemize}

A major advantage of the LATOR mission concept is its independence on both -- the drag-free spacecraft environment and the phase-coherent laser transponding techniques. In fact, LATOR will utilize the photon-counting laser ranging methods and the redundant optical truss provided by the long-baseline optical multi-chanelled interferometer on the ISS. The LATOR experiment is optimized for it's primary science goal -- to measure gravitational deflection of light in the solar gravity to 1 part in 10$^8$ (or at the level of the effects of the second post-Newtonian order of light deflection resulting from gravity's intrinsic non-linearity). There is no technological breakthroughs needed to satisfy the LATOR mission requirements. All the required technologies already exist and most are space-qualified as a part of our on-going interferometry program at JPL (SIM, TPF (Terrestrial Planet Finder), and Palomar Testbed and Keck Interferometers).

Concluding, we point out that the recent progress in relativistic gravity research resulted in a significant tightening of the existing bounds on the PPN parameters obtained at the first post-Newtonian level of accuracy. However, this improvement is not sufficient to lead to groundbreaking tests of fundamental physical laws addressed above. This is especially true, if the cosmological attractor discovered in \cite{damour_nordtvedt2,DPV02,DPV02b} is more robust, time variation in the fine structure constant will be confirmed in other experiments and various GR extensions will demonstrate feasibility of these methods for cosmology and relativistic gravity. The LATOR mission is proposed to directly address the challenges discussed above. 

We shall now discuss the LATOR mission in more details.

\section{Overview of LATOR}
\label{sec:lator_description}

\begin{figure*}[t!]
 \begin{center}
\noindent    
\psfig{figure=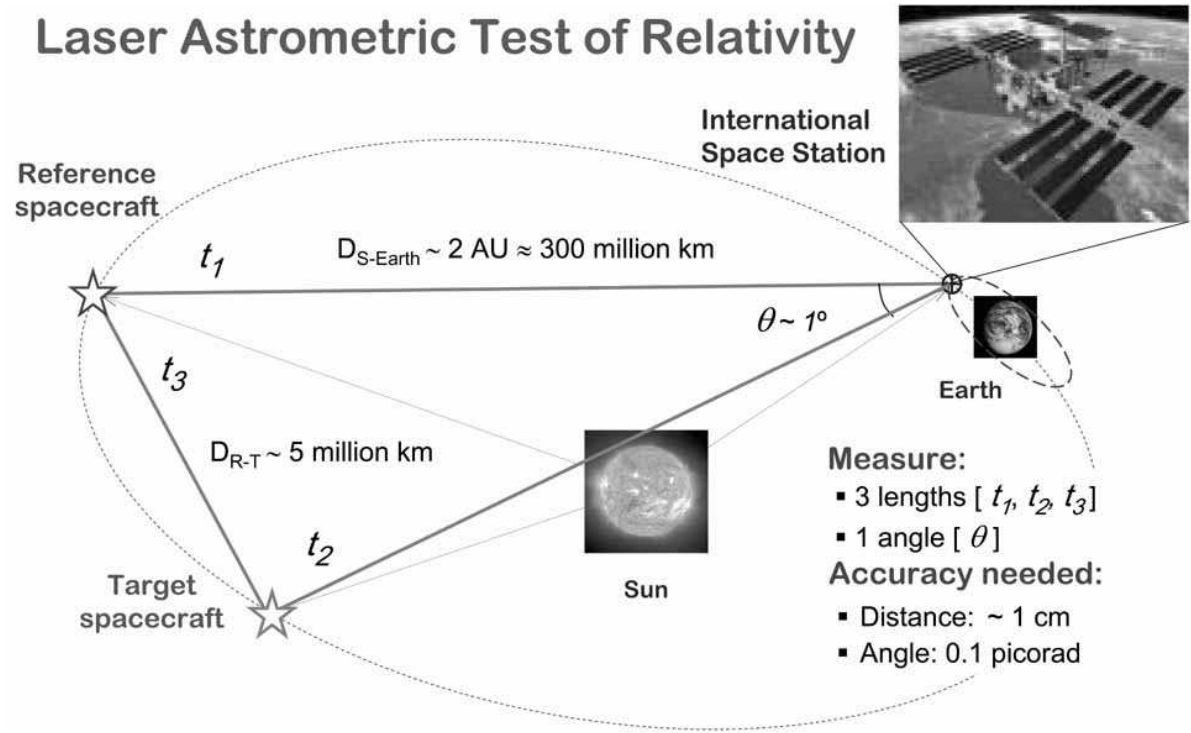,width=158mm}
\end{center}
\vskip -10pt 
  \caption{The overall geometry of the LATOR experiment.  
 \label{fig:lator}}
\end{figure*}

The LATOR experiment uses laser interferometry between two micro-spacecraft (placed in heliocentric orbits, at distances $\sim$ 1 AU from the Sun) whose lines of sight pass close by the Sun to accurately measure deflection of light in the solar gravity. Another component of the experimental design is a long-baseline ($\sim$100 m) multi-channel stellar optical interferometer placed on the ISS. Figure \ref{fig:lator} shows the general concept for the LATOR missions including the mission-related geometry, experiment details  and required accuracies. 

\subsection{LATOR Mission Design}

 The schematic of the LATOR experiment is quite simple (see Fig.~\ref{fig:lator}). Two spacecraft are injected into a solar orbit on the opposite side of the Sun from the Earth. Each spacecraft transmits a laser beam which is detected by a long baseline ($\sim$ 100 m) optical interferometer on the ISS. This interferometer measures the apparent angle between the two spacecraft. In addition, each spacecraft contains laser ranging systems which measure the arms of the triangle formed by the two spacecraft and the ISS. From these measurements, the angle between the two spacecraft viewed from the ISS can be computed using Euclidean geometry. In the absence of gravity, this computed angle will be identical to the apparent angle measured by the interferometer. The difference between the two angles is a measure of the deflection of light by gravity.

As a baseline design for the LATOR orbit, both spacecraft will be launched on the same launch vehicle. Almost immediately after the launch there will be a 30 m/s maneuver that separates the two spacecraft on their 3:2 Earth resonant orbits (see Figure~\ref{fig:lator_traj1}).  The sequence of events that occurs during each observation period will be initiated at the beginning of each orbit of the ISS. It assumed that bore sighting of the spacecraft attitude with the spacecraft transmitters and receivers have already been accomplished. This sequence of operations is focused on establishing the ISS to spacecraft link. The interspacecraft link is assumed to be continuously established after deployment, since the spacecraft never lose line of sight with one another. 

As evident from Figure \ref{fig:lator}, the key element of the LATOR experiment is a redundant geometry optical truss to measure the departure from Euclidean geometry caused by gravity.  The triangle in figure has three independent quantities but three arms are monitored with laser metrology. From three measurements one can calculate the Euclidean value for any angle in this triangle.  In Euclidean geometry these measurements should agree to high accuracy.  This geometric redundancy enables LATOR to measure the departure from Euclidean geometry caused by the solar gravity field to a very high accuracy. The difference in the measured angle and its Euclidean value is the non-Euclidean signal. To avoid having to make absolute measurements, the spacecraft are placed in an orbit where their impact parameters, the distance between the beam and the center of the Sun, vary significantly from $14 R_\odot$ to $1 R_\odot$ over a period of $\sim$ 43 days.

{A version of LATOR with a ground-based receiver was proposed in 1994 and performed under NRA 94-OSS-15 \cite{Shao96}. Due to atmospheric turbulence and seismic vibrations that are not common mode to the receiver optics, a very long baseline interferometer (30 km) was proposed. This interferometer could only measure the differential light deflection to an accuracy of 0.1 $\mu$as, with a spacecraft separation of less than 1 arc minutes.}
In the current version of LATOR, the shortening of the interferometric baseline (as compared to the previously studied version \cite{Shao96}) is achieved solely by going into space to avoid the atmospheric turbulence and Earth's seismic vibrations. On the space station, all vibrations can be made common mode for both ends of the interferometer by coupling them by an external laser truss. This relaxes the constraint on the separation between the spacecraft, allowing it to be as large as few degrees, as seen from the ISS. Additionally, the orbital motion of the ISS provides variability in the interferometer's baseline projection as needed to resolve the fringe ambiguity of the stable laser light detection by an interferometer.

We now outline the basic elements of the LATOR trajectory and optical design.

\subsection{Trajectory -- a 3:2 Earth Resonant Orbit}

\begin{figure}[!t!]
 \begin{center}
\noindent    
\psfig{figure=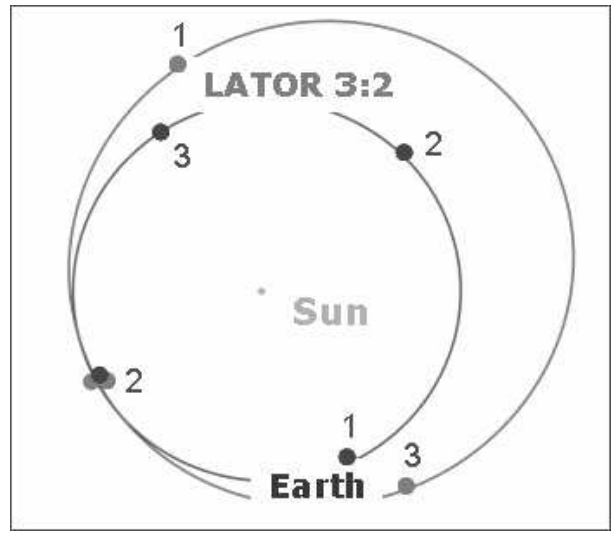,width=80mm}
\end{center}
\vskip -10pt 
  \caption{View from the North Ecliptic of the LATOR spacecraft in a 3:2 resonance. The epoch is taken near the first occultation.
 \label{fig:lator_traj1}}
\end{figure} 
%
\begin{figure*}[!t!]
 \begin{center}
\noindent    
\psfig{figure=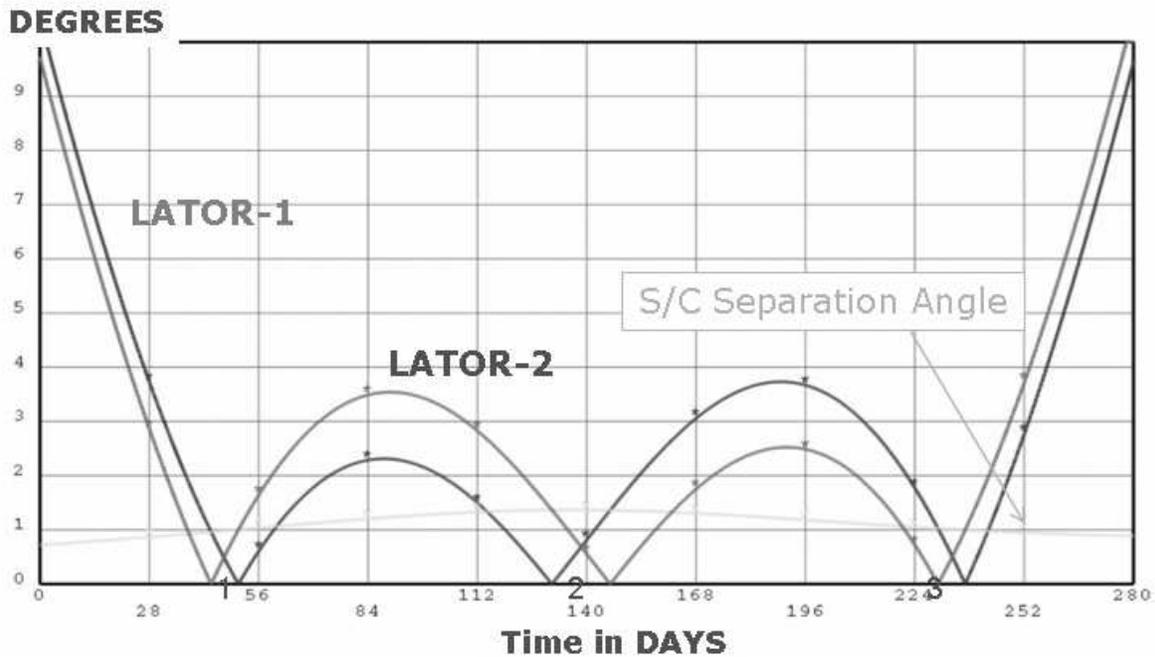,width=158mm}
\end{center}
\vskip -10pt 
  \caption{The Sun-Earth-Probe angle during the period of 3 occultations (two periodic curves) and the angular separation of the spacecraft as seen from the Earth (lower smooth line). Time shown is days from the moment when one of the spacecraft is at 10$^\circ$ distance from the Sun. 
 \label{fig:lator_traj2}}
\end{figure*} 

The objective of the LATOR mission includes placing two spacecraft into a heliocentric orbit with a one year period so that observations may be made when the spacecraft are behind the Sun as viewed from the ISS.  The observations involve the measurement of distance of the two spacecraft using an interferometer on-board the ISS to determine bending of light by the Sun.  The two spacecraft are to be separated by about 1$^\circ$, as viewed from the ISS.  
With the help of the JPL Advanced Project Design Team (Team X), we recently conducted a detailed mission design studies \cite{teamx}. In particular, we analyzed various trajectory options for the deep-space flight segment of LATOR, using both Orbit Determination Program (ODP) and Satellite Orbit Analysis Program (SOAP) --  the two standard JPL navigation software packages. 

One trajectory option would be to use a Venus flyby to place the spacecraft in a 1 yr orbit (perihelion at Venus orbit $\sim$0.73 AU and aphelion $\sim$1.27 AU). One complication of this approach is that the Venus orbit is inclined about 3.4$^\circ$ with respect to the ecliptic and the out-of-plane position of Venus at the time of the flyby determines the orbit inclination \cite{teamx}. The LATOR observations require that the spacecraft pass directly behind the Sun, i.e., with essentially no orbit inclination.  In order to minimize the orbit inclination, the Venus' flyby would need to occur near the time of Venus nodal crossing (i.e., around 7/6/2011). An approach with a type IV trajectory and a single Venus flyby requires a powered Venus flyby with about 500 to 900 m/s.  However, a type I trajectory to Venus with two Venus gravity assists would get LATOR into a desirable 1 year orbit at Earth's opposition.  This option requires no velocity change, called $\Delta v$,  and provides repeated opportunities for the desired science observations. ($\Delta v$ is a desired spacecraft velocity change, which is typically enabled by either the on-board propulsion system or a planetary fly-by.) At the same time this orbit has a short launch window $\sim$17 days which motivated us to look for an alternative.

An good alternative to the double Venus flyby scenario was found when we studied  a possibility of launching LATOR into the orbit with a 3:2 resonance  with the Earth \cite{teamx}. (The 3:2 resonance occurs when the Earth does 3 revolutions around the Sun while the spacecraft does exactly 2 revolutions of a 1.5 year orbit. The exact period of the orbit may vary slightly ($<$1\%) from a 3:2 resonance depending on the time of launch.) For this orbit, in 13 months after the launch, the spacecraft are within $\sim10^\circ$ of the Sun with first occultation occuring in 15 months after launch (See Figures \ref{fig:lator_traj1} and \ref{fig:lator_traj2}).  At this point, LATOR is orbiting at a slower speed than the Earth, but as LATOR approaches its perihelion, its motion in the sky begins to reverse and the spacecraft is again occulted by the Sun 18 months after launch.  As the spacecraft slows down and moves out toward aphelion, its motion in the sky reverses again and it is occulted by the Sun for the third and final time 21 months after launch.  This entire process will again repeat itself in about 3 years after the initial occultation, however, there may be a small maneuver required to allow for more occultations.  Therefore, to allow for more  occultations in the future, there may be a need for an extra few tens of m/s of $\Delta v$. 

The energy required for launch, $C_3$, will vary between $\sim10.6 ~{\rm km}^2/{\rm s}^2 - 11.4 ~{\rm km}^2/{\rm s}^2$ depending on the time of launch, but it is suitable for a Delta II launch vehicle.   The desirable $\sim1^\circ$ spacecraft separation  (as seen from the Earth) is achieved by performing a 30 m/s maneuver after the launch.  This results in the second spacecraft being within $\sim$ 0.6$^\circ$ -- 1.4$^\circ$ separation during the entire period of 3 occultations by the Sun.

Figures \ref{fig:lator_traj1} and \ref{fig:lator_traj2} show the  trajectory  and the occultations in more details.  The first figure is the spacecraft position in the solar system showing the Earth's and LATOR's orbits (in the 3:2 resonance) relative to the Sun.  The epoch of this figure shows the spacecraft passing behind the Sun as viewed from the Earth.  The second figure shows the trajectory when the spacecraft would be within 10$^\circ$ of the Sun as viewed from the Earth.  This period of 280 days will occur once every 3 years, provided the proper maneuvers are performed.  The two similar periodic curves give the Sun-Earth-Probe angles for the 2 spacecraft while the lower smooth curve gives the angular separation of the spacecraft as seen from the Earth. 

The 3:2 Earth resonant orbit provides an almost ideal trajectory for the LATOR spacecraft, specifically i) it imposes no restrictions on the time of launch; ii) with a small propulsion maneuver after the launch, it places the two LATOR spacecraft at the distance of $\leq3.5^\circ$ ($\sim14~R\odot$) for the entire duration of the experiment ($\sim$8 months); iii) it provides three solar conjunctions even during the nominal mission lifetime of 22 months, all within a 7 month period (as opposed to the orbit achieved with two Venus' fly-bys, which offers only 1 conjunction per year); iv) at a cost of an small additional maneuver, it offers a possibility of achieving  small orbital inclinations (to enable measurements at different solar latitudes), and, finally, v) this orbit offers a very slow change in the Sun-Earth-Probe (SEP) angle of $\sim 1R_\odot$ in 4 days. As such, this orbit represents a very attractive choice for the LATOR mission; we intend to further study this trajectory as the baseline option for the mission. In particular, there is an option to have the two spacecraft move in  opposite directions during the solar conjunctions. This option will increase the amount of $\Delta v$ LATOR should carry on-board, but it significantly reduces the experiment's dependence on the accuracy of determination of the solar impact parameter. This particular option is currently being investigated and results will be reported elsewhere. 

We shall now consider the basic elements of the LATOR optical design. 

\subsection{Optical Design}
\label{sec:opt_design}

A single aperture interferometer on the ISS consists of three 20 cm diameter telescopes. One of the telescopes with a very narrow bandwidth laser line filter in front and with an InGAs camera at its focal plane, sensitive to the 1.3 $\mu$m laser light, serves as the acquisition telescope to locate the spacecraft near the Sun.

The second telescope emits the directing beacon to the spacecraft. Both spacecraft are served out of one telescope by a pair of piezo controlled mirrors placed on the focal plane. The properly collimated laser light ($\sim$10W) is injected into the telescope focal plane and deflected in the right direction by the piezo-actuated mirrors. 

The third telescope is the laser light tracking interferometer input aperture which can track both spacecraft at the same time. To eliminate beam walk on the critical elements of this telescope, two piezo-electric X-Y-Z stages are used to move two single-mode fiber tips on a spherical surface while maintaining focus and beam position on the fibers and other optics. Dithering at a few Hz is used to make the alignment to the fibers and the subsequent tracking of the two spacecraft completely automatic. The interferometric tracking telescopes are coupled together by a network of single-mode fibers whose relative length changes are measured internally by a heterodyne metrology system to an accuracy of less than 10 pm.

\begin{figure*}[!t!]
 \begin{center}
\noindent    
\psfig{figure=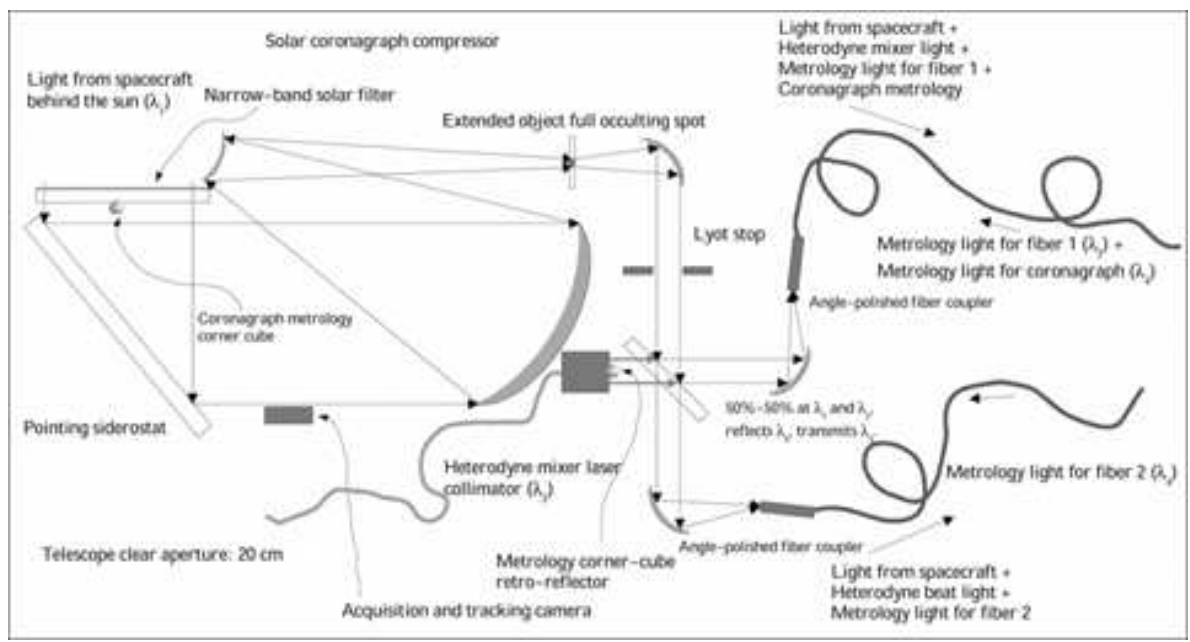,width=157mm}
\end{center}
\vskip -10pt 
  \caption{Basic elements of the LATOR optical design. 
The laser light (together with the solar background) falls onto a full aperture ($\sim 20$cm) narrow band-pass filter with $\sim 10^{-4}$ suppression capabilities and also illuminates the baseline metrology corner cube.  After that, the remaining light is falling onto a steering flat mirror where it will be reflected to an off-axis telescope with no central obscuration (needed for metrology). This is when it enters the solar coronograph compressor by first going through a 1/2 plane focal plane occulter and then coming to a Lyot stop. At the Lyot stop, the background solar light is reduced by a factor of $10^{6}$. The combination of a narrow band-pass filter and coronograph enables the solar luminosity reduction from $V=-26$ to $V=4$ (as measured at the ISS), thus, enabling the LATOR precision observations.
\label{fig:optical_design}}
\end{figure*} 

The spacecraft  are identical in construction and contain a relatively high powered (2 W), stable (2 MHz per hour $\sim$  500 Hz per second), small cavity fiber-amplified laser at 1.3 $\mu$m. Three quarters of the power of this laser is pointed to the Earth through a 20 cm aperture telescope and its phase is tracked by the interferometer. With the available power and the beam divergence, there are enough photons to track the slowly drifting phase of the laser light. The remaining part of the laser power is diverted to another telescope, which points towards the other spacecraft. In addition to the two transmitting telescopes, each spacecraft has two receiving telescopes.  The receiving telescope on the ISS, which points towards the area near the Sun, has laser line filters and a simple knife-edge coronagraph to suppress the Sun light to 1 part in $10^4$ of the light level of the light received from the space station (see Figure \ref{fig:optical_design} for a conceptual design). The receiving telescope that points to the other spacecraft is free of the Sun light filter and the coronagraph.

\begin{figure}[h!]
 \begin{center}
\noindent  \vskip -5pt   
\psfig{figure=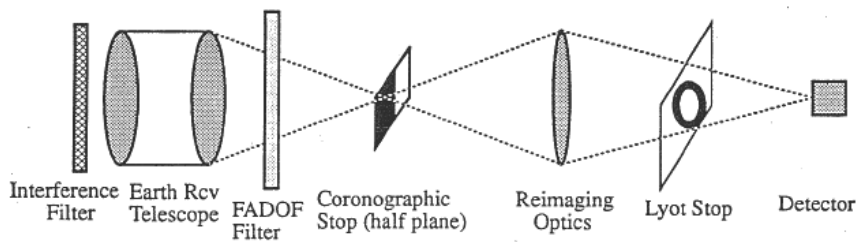,width=115mm}
\end{center}
\vskip -15pt 
  \caption{LATOR coronograph system.
 \label{fig:coronograph}}
\end{figure} 

In order to have adequate rejection of the solar background surrounding the laser uplink the ISS and spacecraft optical systems must include a coronagraph. Figure \ref{fig:coronograph} shows a schematic of the coronagraph. A 20~cm telescope forms an image on the chronographic stop. This stop consists of a knife-edge mask placed 6 arcseconds beyond the solar limb. The transmitted light is then reimaged onto a Lyot stop, which transmits 88\% of the incident intensity. Finally, the light is
reimaged onto the tracking detector. Note that, in addition a combination of a wideband interference filter and a narrow band Faraday anomalous dispersion optical filter (FADOF) (see Fig.~\ref{fig:coronograph}) will be used to reject light outside a 0.005 nm band around the laser line. Thus, for ISS-SC receiver/transmitter, the incoming signal will subdivided with one portion going to a high bandwidth detector and the other to an acquisition and tracking CCD array (see Fig.~\ref{fig:optical_design}). Using a $64 \times 64$ CCD array with pixels sized to a diffraction limited spot, this array will have a 5 arcmin field of view which is greater than the pointing knowledge of the attitude control system and the point ahead angle (40 arcsec). After acquisition of the ISS beacon, a $2\times 2$ element subarray of the CCD will be used as a quad cell to control the ISS-SC two axis steering mirror. This pointing mirror is common to both the receiver and transmitter channel to minimize misalignments between the two optical systems due to thermal variations. The pointing mirror will have 10 arcminute throw and a pointing accuracy of 0.5 arcsec which will enable placement of the uplink signal on the high bandwidth detector. Similar design elements will be implemented in the other optical packages.

Our preliminary analysis indicates that LATOR will achieve a significant stray light rejection, even observing at the solar limb. In fact, the flux from the solar surface may be minimized by a factor of $10^4$. In addition to decreasing the stray solar radiation, the coronograph will decrease the transmission of the laser signal by 78\% (for a signal 12 arcsec from limb) due to coronographic transmission and broadening of the point spread function. At these levels of solar rejection, it is possible for the spectral filter to reject enough starlight to acquire the laser beacon (even at the $\sim$~fW level). It is interesting that without the coronograph, the stray light from the Sun, decreases proportionally to the distance from the limb, but with the use of the coronograph, it decreases as a square of the distance from the limb. The results of our simulations will be reported in the subsequent publications.

In addition to the four telescopes they carry, the spacecraft also carry a tiny (2.5 cm) telescope with a CCD camera. This telescope is used to initially point the spacecraft directly towards the Sun so that their signal may be seen at the space station. One more of these small telescopes may also be installed at right angles to the first one to determine the spacecraft attitude using known, bright stars. The receiving telescope looking towards the other spacecraft may be used for this purpose part of the time, reducing hardware complexity. Star trackers with this construction have been demonstrated many years ago and they are readily available. A small RF transponder with an omni-directional antenna is also included in the instrument package to track the spacecraft while they are on their way to assume the orbital position needed for the experiment. 

The LATOR experiment has a number of advantages over techniques which use radio waves to measure gravitational light deflection. Advances in optical communications technology, allow low bandwidth telecommunications with the LATOR spacecraft without having to deploy high gain radio antennae needed to communicate through the solar corona. The use of the monochromatic light enables the observation of the spacecraft almost at the limb of the Sun, as seen from the ISS. The use of narrowband filters, coronagraph optics and heterodyne detection will suppress background light to a level where the solar background is no longer the dominant noise source. In addition, the short wavelength allows much more efficient links with smaller apertures, thereby eliminating the need for a deployable antenna. Finally, the use of the ISS will allow conducting the test above the Earth's atmosphere -- the major source of astrometric noise for any ground based interferometer. This fact justifies LATOR as a space mission.

\begin{figure*}[t!]
 \begin{center}
\epsfig{figure=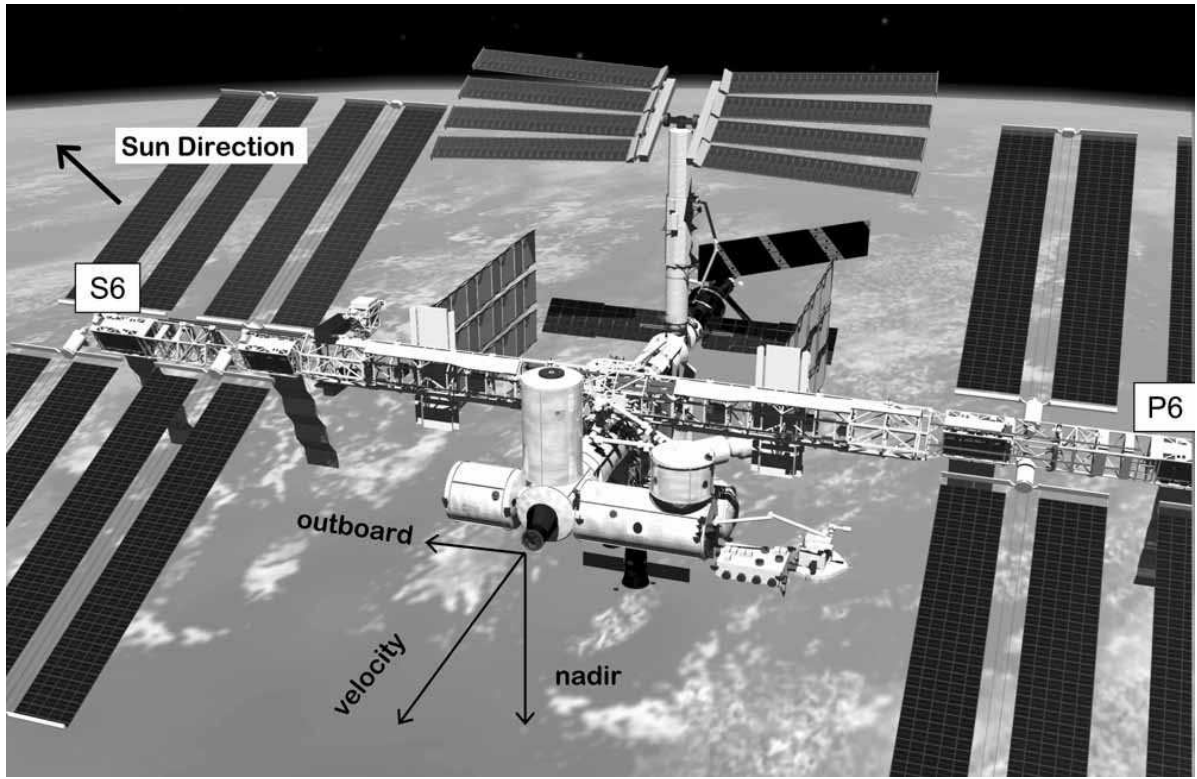,width=158mm}
\end{center}
\vskip -10pt 
  \caption{Location of the LATOR interferometer on the ISS. To utilize the inherent ISS sun-tracking capability, the LATOR optical packages will be located on the outboard truss segments P6 and S6 outwards. 
 \label{fig:iss_config}}
\end{figure*} 

\subsection{Interferometer on the ISS}
\label{sec:intenferometer}

Interferometer will be formed by the two packages containing the entire optical elements discussed above. Each such assembly will have dimensions of 0.8~m~$\times$ 0.8~m~$\times$~0.8~m and will weight $\sim$70 kg. The location of these packages on the ISS and their integration with the ISS's power, communication and attitude control system are discussed below:

\begin{itemize}
\item Two LATOR optical packages will be physically located and integrated with the ISS infrastructure (see Fig.~\ref{fig:iss_config}). The location will enable the straight-line separation between the two fiducials forming the $\sim$100 m baseline and will provide a clear line-of-site (LOS) path between the two transponders during the observation periods.  

\item The optical packages will be located on the ISS structure to maximize the inherent ISS sun-tracking capability.  The telescopes will need to point towards the Sun during each observing period; thus, by locating these payloads on the ISS outboard truss segments (P6 and S6 outwards), a limited degree of automatic sun-tracking capability is afforded by the alpha-gimbals on the ISS.

\item The minimum unobstructed LOS time duration between each telescope on the ISS and the telescope/transponders and their respective spacecraft shall be 58 minutes per the 92 min orbit of the ISS.  This is the minimum time requirement to make useful scientific measurements of the gravitational phenomena.
 
\item	The pointing error of each instrument will be of the order of 1~$\mu$rad for control,  1~$\mu$rad for knowledge, with a stability of 0.1~$\mu$rad/sec. Each transponder will have real-time ISS attitude data with an accuracy $\sim$1 $\mu$rad (absolute) and 0.1~$\mu$rad/s (rate of change), provided by the standard GPS link available on the ISS.

\item Both optical packages will be powered by the ISS electrical power system (EPS). Each assembly will require $\sim$150 W of electrical power while making measurements and $\sim$60 W during survival and standby modes.

\item	Each assembly will generate maximum 10 Kbps of data for downlink to the ground. The ISS will downlink the stored  transponder data to the ground with frequency once per every 8 hours while active measurements are being made and 1 time every day while the instruments are in standby mode.
\end{itemize}

The LATOR mission will utilize several technology solutions that recently became available. In particular, signal acquisition on the solar background will be done with a full-aperture narrow band-pass filter and coronagraph discussed above. The issue of the ISS's extended structure vibrations will be solved by integrating $\mu$-g accelerometers in the optical packages. The accelerometers will provide information needed to improve the attitude information of each telescope and fiducial within the interferometer. (The use of  accelerometers was first devised for SIM, but at the end their application was not necessary; however, the  Keck Interferometer uses accelerometers extensively to address the issue of the seismic noise contribution in the interferometric astrometry observations.) Finally, the problem of monochromatic fringe ambiguity that complicated the design of the previous version of the experiment \cite{yu94,Shao96} and led to the use of variable baselines lengths -- is not an issue for LATOR. The orbit of the ISS  provides at least a 30\% variation in the baseline projection which  solves this problem.  

We shall now consider the LATOR preliminary astrometric error budget.

\section{LATOR Astrometric Error Budget}
\label{sec:error_bud}

The goal of measuring deflection of light in solar gravity with accuracy of one part in $10^{8}$ requires serious consideration of systematic errors. This work requires a significant effort to properly identify the entire set of factors that may influence the accuracy at this level. Fortunately, we initiated this process being aided with experience of successful development of a number of instruments that require similar technology and a comparable level of accuracy, notably SIM, TPF, Keck and Palomar Testbed Interferometers. This experience comes with understanding various constituents of the error budget, expertize in developing appropriate instrument models; it is also supported by the extensive verification of the expected  performance with the set of instrumental test-beds (designed and built solely for this purpose) and existing flight hardware. Details of the LATOR error budget are still being developed (especially those of the second order) and will be published elsewhere, when fully analyzed. Here we discuss a preliminary astrometric error budget for the LATOR experiment and  present design considerations that enable the desirable instrument performance. 

It is convenient to present error sources in three broad categories: i)~the ones that are related to mission architecture, ii) those that are external to the triangle and have an astrophysical origin, and iii) those that  originated within the instrument itself. 
Typical mission-related errors are those that result from the uncertainties in the orbits of the spacecraft and the ISS,  chosen mission design and observing scenario and, in general, those errors that result from the  geometry of the experiment and affect the range and angle determination. 
The astrophysical errors are those that are external to the instrument and are due to various phenomena that influence both  mission planning and observing scenario. Such errors are due to the gravity effects of planets and largest asteroids,  unmodeled motion of the fiducial stars, optical properties of the Sun and solar corona near the limb, etc. 
Example of the instrumental errors include effects of the long term laser stability, errors in pointing of the laser beams, instrumental drifts and other systematic and random errors originating within the instrument itself. (By instrument we understand the experimental hardware situated at all three vortices of the triangle.)

We will discuss the contributions of these factors below, but first we shall introduce the model for the LATOR astrometric observable.

\subsection{Observational Model} 
To develop the error budget we utilize a simple model to capture all error sources and their individual impact on the mission performance. (A more detailed model to the second order in gravitational effects is available and is being used in simulations to verify the expected mission performance.) 

To the first order in gravitational constant,  the light paths, $\ell_{ij}$, between the three vortices of the triangle  may be given as follows (i.e. a usual Shapiro time delay relation)
\begin{equation}
\ell_{ij}=r_{ij}+(1+\gamma)\mu_\odot\ln[\frac{r_i+r_j+r_{ij}}{r_i+r_j-r_{ij}}], ~~~~ \mathbf{r}_{ij}=\mathbf{r}_j-\mathbf{r}_i,
\label{eq:path}
\end{equation}
\noindent where $\mathbf{r}_i$ is the barycentric Euclidian position to one of the three vortices, $i,j\in\{1,3\}$ ($i=3$ is for the ISS), with $r_i=|\mathbf{r}_i|$, being its distance, and $\mu_\odot=GM/c^2$ is the solar gravitational radius. To a similar accuracy, the interferometric delay, $ d_{j}$, for a laser source $j$ has the following approximate form (i.e. differenced Shapiro time delay for the two telescopes separated by an interferometric baseline, $\mathbf{b}$, or $d_j=\ell_{j3}(\mathbf{r}_3)-\ell_{j3}(\mathbf{r}_3+\mathbf{b})$):
{}
\begin{equation}
d_j\simeq(\mathbf{b}\cdot\mathbf{n}_{j3})-(1+\gamma)\mu_\odot 
\frac{2r_jr_3}{r_3+r_j}\frac{\mathbf{b}\cdot(\mathbf{n}_3-\mathbf{n}_{j3})}{p_j^2},
\label{eq:delay}
\end{equation} 
\noindent where $p_j$ is the solar impact parameter for source $j$. Both expressions Eqs.(\ref{eq:path}) and (\ref{eq:delay})  require some additional transformations to keep only the terms with a similar order. 
The entire LATOR model would have to account for a whole range of other effects, including terms due to gravitational multipoles, second order deflection, angular momentum contribution, and etc. This interesting work had being initiated and the corresponding results will be reported elsewhere. Therefore, below we shall comment only on the conceptual formulation of the LATOR observables. 

The range observations Eq.(\ref{eq:path}) may be used to measure any angle between the three fiducials in the triangle. However, for observations in the solar gravity field, measuring the lengths do not give you a complete information to determine the angles, and some extra information is needed. This information is the mass of the Sun, and, at least one of the impact parameters. Nevertheless, noting that the paths $\underline{\ell}_{ij}$ correspond to the sides of the connected, but gravitationally distorted triangle, one can write $\underline{\ell}_{12}+\underline{\ell}_{23}+\underline{\ell}_{31}=0$, where $\underline{\ell}_{ij}$ is the null geodesic path for light moving between the two points $i$ and $j$. This leads to the expression for the angle between the spacecraft $\cos(\widehat{\underline{\ell}_{31}\underline{\ell}_{32}})=\cos\delta_r =(\ell_{32}^2+\ell_{31}^2-\ell_{12}^2)/(2 \ell_{32} \ell_{31})$. Expression for $\cos\delta_r$ will have both Euclidian and gravitational contributions; their detailed form will not significantly contribute to the discussion below and, thus, it is outside the scope of this paper.

The astrometric observations Eq.(\ref{eq:delay}) will be used to obtain another measurement of the same angle between the two spacecraft. The LATOR interferometer will perform differential observations between the two sources of laser light, measuring the differential delay $\Delta d_{12}=d_2-d_1$ to the accuracy of less than $5$ pm (see below). For the appropriate choice of the baseline orientation, one can present the angle between the two sources of laser light as  $\cos(\widehat{\underline{\ell}_{31}\underline{\ell}_{32}})=\cos\delta_a=1-\Delta d_{12}^2/2b^2$. This expression would have both Euclidian and gravity contributions which are not discussed in detail in this paper.

The two sets of observations obtained by laser ranging and astrometric interferometry form the complete set of LATOR observables. Conceptually,  the LATOR astrometric measurement $\delta_d$ of the gravitational deflection of light may be modeled as
\begin{equation}
\delta_d = \delta_r-\delta_a = c_1\Big(\frac{1}{p} - \frac{1}{p+\Delta p}\Big)+c_2\Big(\frac{1}{p^2} - \frac{1}{(p+\Delta p)^2}\Big),
\end{equation}
 
\noindent where  $\delta_r$ is the angle computed from the range information,  $\delta_a$ is the angle measured astrometrically by the interferometer.
$p$ is the impact parameter of the spacecraft closer to the Sun and $\Delta p$ is the difference between the two impact parameters. $c_1$ and $c_2$ are the first and second order terms in the gravitational deflection and are the quantities of interest. Three such measurements are made to simultaneously solve for these constants together with the impact parameter. 
The temporal evolution of the entire triangle structure will produce another set of observables $\delta_v=d\delta_d/dt=(\partial \delta_d/\partial p)dp/dt$ which will be used to process the data (similar to the method highlighted in \cite{iess_etal_1999, cassini_and}). A fully relativistic model for this additional independent observable, including the contributions of range and angle rates, is being currently developed.  

The error budget is subdivided into three components -- range and interferometer measurements, and spacecraft stability which are described below.

\subsubsection{Range Measurement:} 
 
This component describes the angular errors due to uncertainties in the distance between the spacecraft and the ISS as they determined by laser ranging. The angular error due to an ISS-to-spacecraft ranging error is

\begin{equation}
\Delta _{D}= \frac{\rho}{D^2} \delta D
\end{equation}
 
\noindent where  $\delta D$ is the ranging accuracy, $\rho$ is the distance between the spacecraft ($\rho\simeq1^\circ =5.22~\times 10^6$~ km), and $D$ is the ISS-to-spacecraft distance ($D\simeq 2 {\rm AU} = 3 \times 10^{11}$~m). We have allocated a 60 cm range uncertainty for each of the laser links, which for spacecraft separated by 1$^\circ$, results in an angular uncertainty of 0.035 prad (i.e. $\sigma_d=3.5$~pm). 

The angular uncertainty due to an inter-spacecraft ranging error, $\Delta \rho$, is
  
\begin{equation}
\Delta \rho= \frac{\delta {\rho}}{D} 
\end{equation}

\noindent The experiment will require a spacecraft-to-spacecraft laser ranging accuracy of 1 cm, resulting in an angular error of 0.03 prad (i.e. $\sigma_d=3$~pm). The total error budget for the laser ranging distance measurements is $4.6$~pm.

\subsubsection{Interferometer Measurement:} 
Experiment uncertainties in the ISS interferometer measurement contribute additional terms to the overall error budget.   The baseline design for the instrument is a 100~m baseline with a 10 s integration time. Based on theoretical predictions for narrow angle measurements, this configuration will result in an angular error of 0.025 prad (i.e. $\sigma_d=2.5$~pm), limited by the long-term instrument systematics. 

The current requirement for systematic errors in the instrument has been set at 0.05 prad. This corresponds to measurement of the laser fringe phase to 1 part in $2.5\times10^5$ ($\lambda =1.3 ~\mu$m, $b = 100$~m). This term includes errors in the metrology and fringe detection of the interferometer, as well as the effect of photon noise. 

\subsubsection{Spacecraft Orbit Stability:}
 
In order to determine the first and second order terms of gravitational deflection, LATOR will make a number of measurements at different spacecraft separations and various impact parameters. The baseline experiment calls for three measurements to be made in order to solve for the impact parameter together with  deflection terms, $c_1$ and $c_2$. During the period between measurements, it is assumed that the impact parameter is known. An error in this assumption will cause an equivalent error in the computation of the deflection term which is given by
 
\begin{equation}
\Delta \delta_{d}= 
c_1 \Big(\frac{1}{p^2}-\frac{1}{(p+\Delta p)^2}\Big) \Delta {p},
\end{equation}

\noindent where $\Delta {p}$ is the uncertainty in the impact parameter. For simplicity we approximate $c_1 = (4GM/c^2R_\odot)R_\odot b \simeq 8\times 10^{-6} R_\odot b$, which is for the gravitational deflection at the limb of the Sun ($p=R_\odot\simeq0.26^\circ$). We require that the spacecraft be stable to 0.05 mas, which corresponds to a drift in the transverse distance of 20 m and results in an angular error of 0.04 prad (i.e. $\sigma_d=4$~pm).

\subsubsection{ISS Orbit Stability:} 

In addition to the spacecraft error, the ISS's orbital error will also produce contribution to the angular measurement. Most of the errors on the ISS can be made common-mode; therefore, their influence on the differential astrometry with LATOR interferometer will be either negligible  or it will be small and well modeled. However, there are some errors that would still produce measurable contribution to the differential delay, if not properly addressed; notably, the accuracy of the ISS orbit.  The current mission design calls for an enhancement of the ISS orbit solution by utilizing GPS receivers at the location of each optical (see Section~\ref{sec:intenferometer}). This will also help to address the issue of the extended structure low-frequency vibrations of the ISS. As we mentioned above, the effect of these vibrations will be addressed by using $\mu$-g level accelerometers, that will be integrated within both optical packages on the ISS. A combination of the GPS receivers and $\mu$-g accelerometers will provide information needed to improve the ISS attitude information; this improvement will be done for each cornercube fiducial (needed for the interferometric baseline determination). Our current error budget for the differential observations with the LATOR interferometer allocates $\sim 2.7 $~pm of error in 100 s of integration for the uncertainty in the ISS orbit, its attitude and the extended structure vibrations.

\subsection{Mission Errors}

In the current design, the LATOR experiment requires that the location of one of the spacecraft with respect to the Sun is known to within 20~m over the duration of the each observing session or $\sim $ 92 min. The major perturbation to the spacecraft trajectory is from local spacecraft disturbances, such as gas leaks for thruster valves and solar radiation pressure. The spacecraft can be designed to eliminate spacecraft errors leaving solar radiation pressure as the major source for the position noise. Other disturbances such as solar wind, magnetic fields, cosmic rays, etc. have been identified and are at least three orders of magnitude lower than solar radiation pressure. 
 

\subsubsection{Direct Solar Radiation Pressure:}
\label{sec:solar_plasma}

There is an exchange of momentum when solar photons impact the
spacecraft and are either absorbed or reflected.  Models for 
this solar pressure effect are usually developed before a mission is launched. The models take into account various parts of the spacecraft exposed to solar radiation; they compute the acceleration directed away from the Sun as a function of spacecraft orientation and solar distance: 
\begin{equation}
a_{\tt s.p.}(r)=\frac{2 f_\odot A }{c~m}
\frac{ \cos\theta(r)}{r^2},
 \label{eq:srp}
\end{equation}
\noindent where $f_\odot=1367 ~{\rm W/m}^{2}$(AU)$^2$ is the 
(effective-temperature Stefan-Boltzmann) 
``solar radiation constant'' at 1 AU from the Sun and $A$ is the effective  size of the craft as seen by the Sun.   $\theta$ is  the angle between the axis of the antenna and the direction of the Sun, $c$ is the speed of light, $M$ is the mass of the spacecraft, and $r$ is the distance from  the Sun to the spacecraft in AU.   
For expected spacecraft values of $A= 0.3 ~{\rm m}^2$ and $m = 100 ~{\rm kg}$, gives an acceleration of $a_{\tt s.p.} \simeq 2.7 \times 10^{-8}~{\rm  m}/{\rm s}^2$ at $r=1$~AU from the Sun.

This acceleration will produce an unmodeled force, which ultimately may result in the error in the radial position of the spacecraft. Over a time $t$ this error is 
$\delta r = \frac{\delta a}{2}t^2$ 
where  ${\delta a}$ is the unmodeled acceleration. In turn, this error would lead to a transverse position
error of $\delta x = \frac{\delta a}{4}n t^3$ 
where $n$ is the spacecraft velocity about the Sun, $n\sim 2 \times 10^{-7}$~rad/s.
If the effect of solar radiation pressure were completely unmodeled, over a period of 20 days, the transverse position error due to solar pressure would be $\sim$~7 km. Consequently, it is necessary to use the laser ranging information to predict the transverse position of the spacecraft. With a 60 cm ranging knowledge, the transverse position uncertainty over a period of 20 days would only be 10 cm. 

The laser ranging information will be used to solve for the slowly varying changes in the solar pressure leaving the random fluctuations of the solar pressure as the dominant source of position error. In the case when random fluctuations corresponding to 1\% {\small RMS} of the total solar radiation pressure, the spacecraft wanders $\sim$~1~m in the radial direction and 8~cm in the transverse direction in a day.
{}
The use of a redundant optical truss offers an excellent alternative to an ultra-precise orbit determination. This feature also makes LATOR insensitive to spacecraft buffeting from solar wind and solar radiation pressure. This is why, as opposed to other gravitational missions, there is no need for a drag-free spacecraft to enable the high accuracy of the LATOR experiment. 

\subsubsection{Effect of the Solar Corona:}
\label{sec:solar_corona}

For the navigation purposes, both LATOR spacecraft will be equipped with X-band transponders with both Doppler and range capabilities. The electron density and density gradient in the solar atmosphere will influence the propagation of radio waves through the medium. So, both range and Doppler observations at X-band will affected by the electron density in the interplanetary medium and outer solar corona. This would result in the spacecraft not being able to communicate with the ground when the impact parameter will be less than $\sim 2.5R_\odot$.  This is why the current mission plan includes provision that the radio-navigation will not be conducted for the solar impact parameters smaller than $2.5R_\odot$. The most of the important navigational, instrumental and experimental information will be stored on-board until the time of clear communication with Deep Space Network. This is when the mission will step-up to its full potential by enabling communication from these extremely small distances from the Sun by utilizing its optical communication capabilities to enable a high precision  spacecraft navigation. 

The use of optical wavelengths offer a significant advantage for the spacecraft communication in the solar system as opposed to the microwave radiation -- the current navigation standard. Such a choice makes the deep space communication effectively free from the solar corona noise. Indeed, the solar plasma effects on wave propagation decrease as $\lambda^2$ and there is a factor of $10^{10}$ reduction in the solar plasma optical path fluctuations by simply moving from the S-band microwave signal $\lambda=10$~cm ($f= 3$~GHz) to the optical wavelengths of $\lambda\sim1 ~\mu$m ($f= 300$ THz). This $10^{10}$ reduction of the dispersive media effects offers tremendous gain in the quality of both spacecraft navigation (increased pointing precision and timing) and deep space communication (very high data transmission rates). LATOR will utilize design capable of rejecting background solar noise in combination with optical  wavelengths for precision navigation; this combination will lead to a significant reduction of the solar corona effect, making its contribution harmless to the mission.

\subsection{Astrophysical Errors}

Physical phenomena of an astrophysical origin that are external to the LATOR triangle, but do not affect the mission navigation accuracy, are treated as the sources of astrophysical errors.  These errors would  nominally influence both the mission planning and observing scenario, they would be due to non-stationary behavior of the gravity field in the solar system (gravity effects of planets and  asteroids),  unmodeled motion of the fiducial stars, optical properties of the Sun and solar corona near the limb, the properties of the solar surface and etc. We will discuss these sources in some detail. 

\subsubsection{Knowledge of Solar Interior:}
Laser ranging between the ISS and  spacecraft will be used to measure the orbits of the flight segments with $\sim$1 cm accuracy. This implies that the solar impact parameter should be measured to $1.5\times10^{-9}$, a scaling error for the measurement of parameter $\gamma$, but an insignificant error for the other measurements. 

Along with the impact parameter, other solar parameters such as its mass, angular momentum and quadrapole moment must be also known (or will be solved for directly from the data). 
The LATOR instrument may actually be used to gain additional knowledge on the Sun by observing its surface with a Doppler imager. This information may than be used to study the propagation of the sound waves through the solar interior. The resulted data may be used to bootstrap the gravitational solution for the solar oblateness and the higher spherical multipoles of the solar interior.  The instrumental implication of this possibility are currently being investigated and, if feasible, it may be included for the mission proposal. 

\subsubsection{Solar System Gravity:}
Since the solar system is not static and the spacecraft are in the orbits
around the Sun, many large solar system bodies, such as the Sun itself, planets, asteroids, and even the galaxy, would have a significant effect on the measurement of $\gamma$ at the eighth decimal place.  Fortunately the ephemerides for the solar system objects are known to sufficient accuracy and the motion of the solar system about the galactic center is sufficiently smooth during the 92 min of each observing session.  Earth orbit crossing asteroids may cause a significant disturbance if they come within $\sim$10,000 km of one of the arms of the triangle. The relativity measurement may either have to be delayed or conducted with a slightly higher sampling rate if one of these are nearby. The change in the first order relativistic time delay due to other bodies  in the solar system has to be known to $\sim 10^{-9}$ of the effect from the Sun. The final observational model would have to account for the effects due to all the major solar system bodies. A similar theoretical and algorithmic work is currently being conducted for both SIM and GAIA interferometers and may well be used for this mission.

\subsection {Instrument Errors}

In our design considerations we address two types of instrumental errors, namely the offset and scale errors. Thus, in some cases, when a measured value has a systematic offset of a few pm, there are may be instrumental errors that lead to further offset errors.  There are many sources of offset (additive)  errors caused by imperfect optics or imperfectly aligned optics at the pm level; there also many sources for scale errors. We take a comfort in the fact that, for the space-based stellar interferometry, we have an ongoing technology program at JPL;  not only this program has already demonstrated metrology accurate to a sub-pm level, but has also identified a number of the error sources and developed methods to either eliminate them or to minimize their effect at the required level.

The second type of error is a scale error. For instance, in order to measure $\gamma$ to one part in  $10^{8}$ the laser frequency also must be stable to at least to $10^{-8}$ long term; the lower accuracy would result in a scale error. The measurement strategy adopted for LATOR would require the laser stability to only $\sim$1\% to achieve accuracy needed to measure the second order gravity effect. Absolute laser frequency must be known to $10^{-9}$ in order for the scaling error to be negligible. Similarly robust solutions were developed to address the effects of other known sources of scale errors. 

There is a considerable effort currently underway at JPL to evaluate a number of potential errors sources for the LATOR mission, to understand their properties and establish methods to mitigate their contributions. (A careful strategy is needed to isolate the instrumental effects of the second order of smallness; however, our experience with SIM \cite{mct, Turyshev01_1, Turyshev01_2} is critical in helping us to properly capture their contribution in the instrument models.)  The work is ongoing, this is why the discussion below serves for illustration purposes only. We intend to publish the corresponding analysis and simulations in the subsequent publications.

\subsubsection{Optical Performance:}
 
The laser interferometers use $\sim$2W lasers and $\sim$20 cm optics for transmitting the light between spacecraft. Solid state lasers with single frequency operation are readily available and are relatively inexpensive.   For SNR purposes we assume the lasers are ideal monochromatic sources (with $\lambda = 1.3~ \mu$m). For simplicity we assume the lengths being measured are 2AU = $3\times 10^8$ km. The beam spread is estimated as $\sim 1~\mu$m/20~cm = 5 $\mu$rad (1 arcsec). The beam at the receiver is $\sim$1,500 km in diameter, a 20 cm receiver will detect $1.71 \times 10^2$ photons/s assuming 50\% q.e. detectors. Given the properties of the CCD array it takes about 10 s to reach the desirable SNR of $\sim2000$ targeted for the detection of the second order effects. In other words, a 5 pm resolution needed for a measurement of the PPN parameter $\gamma$ to the accuracy of one part in $\sim10^{8}$ is possible with $\approx10$~s of integration.

As a result, the LATOR experiment will be capable of measuring the angle between the two spacecraft to $\sim0.05$~prad, which allows light deflection due to gravitational effects to be measured to one part in $10^8$. Measurements with this accuracy will lead to a better understanding of gravitational and relativistic physics. In particular, with LATOR, measurements of the first order gravitational deflection will be improved by a factor of 3,000. LATOR will also be capable of distinguishing between first order ($\propto G$) and second order ($\propto G^2$) effects. All effects, including the first and second order deflections, as well as the frame dragging component of gravitational deflection and the quadrupole deflection will be measured astrometrically.  

In our analysis we have considered various potential sources of systematic error. 
This information translates to the expected accuracy of determination of the differential interferometric delay of $\sim \pm5.4$ pm, which enables measurement of PPN parameter $\gamma$ to accuracy of  
$\gamma-1 = \pm 0.9  \times 10^{-8}.$
This expected instrumental accuracy is clearly a very significant improvement compared to other currently available techniques. This analysis serves as the strongest experimental motivation to  conduct the LATOR experiment.  

\subsection{\label{sec:expect_accuracy}Expected Measurement Accuracy}

Here we summarize our estimates of the expected accuracy in measurement of the relativistic parameters of interest.
As Table \ref{tab:eff} suggests, the first order effect of light deflection in the solar gravity caused by the solar mass monopole is $\alpha_1=1.75$ arcsec; this value corresponds to an interferometric delay of $d\simeq b\alpha_1\approx0.85$~mm on a $b=100$~m baseline. Using laser interferometry, we currently able to measure  distances with an accuracy (not just precision but accuracy) of $\leq$~1~pm. In principle, the 0.85 mm gravitational delay can be measured with $10^{-9}$ accuracy versus $10^{-5}$ available with current techniques. However, we use a conservative estimate  of 10 pm for the accuracy of the delay which would lead to a single measurement of $\gamma$ accurate to 1 part in $10^{8}$ (rather than 1 part in $10^{9}$), which would be already a factor of 3,000 accuracy improvement when compared to the recent Cassini result \cite{cassini_ber}. 

Furthermore, we have targeted an overall measurement accuracy of 10 pm per measurement, which for $b=100$~m this translates to the accuracy of 0.1 prad $\simeq 0.02 ~\mu$as. With 4 measurements per observation, this yields an accuracy of $\sim5.8\times 10^{-9}$ for the first order term.
The second order light deflection is approximately 1700 pm and with 10 pm accuracy and the adopted measurement strategy it could be measured with accuracy of $\sim2\times 10^{-3}$, including first ever measurement of the PPN parameter $\delta$.  The frame dragging effect would be measured with $\sim 1\times10^{-2}$ accuracy and the solar quadrupole moment (using the theoretical value of the solar quadrupole moment $J_2\simeq10^{-7}$) can be modestly measured to 1 part in 20, all with respectable signal to noise ratios.


\section{Conclusions}
\label{sec:conc}

LATOR mission is the 21st century version of Michelson-Morley experiment searching for a  cosmologically evolved scalar field in the solar system. This mission aims to carry out a test of the curvature of the solar system's gravity  field with an accuracy better than 1 part in 10$^{8}$. In spite of the previous space missions exploiting radio waves for tracking the spacecraft, this mission manifests an actual breakthrough in the relativistic gravity experiments as it allows to take full advantage of the optical techniques that recently became available.  LATOR has a number of advantages over techniques that use radio waves to measure gravitational light deflection. Thus, optical technologies allow low bandwidth telecommunications with the LATOR spacecraft. The use of the monochromatic light enables the observation of the spacecraft at the limb of the Sun. The use of narrowband filters, coronagraph optics and heterodyne detection will suppress background light to a level where the solar background is no longer the dominant noise source. The short wavelength allows much more efficient links with smaller apertures, thereby eliminating the need for a deployable antenna. Finally, the use of the ISS enables the test above the Earth's atmosphere -- the major source of astrometric noise for any ground based interferometer. This fact justifies LATOR as a space mission.

The LATOR experiment technologically is a very sound concept; all technologies that are needed for its success have been already demonstrated as a part of the JPL's interferometry program.  
	The LATOR experiment does not need a drag-free system, but uses a geometric redundant optical truss to achieve a very precise determination of the interplanetary distances between the two micro-spacecraft and a beacon station on the ISS. The interest of the approach is to take advantage of the existing space-qualified optical technologies leading to an outstanding performance in a reasonable mission development time.  The  availability of the ISS makes this mission concept realizable in the very near future; the current mission concept calls for a launch as early as in 2011 at a cost of a NASA MIDEX mission.   

Our next steps will be to perform studies of the optimal trajectory configuration and mission design including the launch vehicle choice trade studies. Our analysis will concentrate on the thermal design of the instrument, analysis of the launch options and configuration, estimates of on-board power and weight requirements as well as analysis of optics and vibration contamination for the interferometer.  We also plan to  develop an end-to-end mission simulation, including detailed mission error budget. 

LATOR will lead to very robust advances in the tests of fundamental physics: it could discover a violation or extension of GR, or reveal the presence of an additional long range interaction in the physical law.  There are no analogs to the LATOR experiment; it is unique and is a natural culmination of solar system gravity experiments. 

The work described here was carried out at the Jet Propulsion Laboratory, California Institute of Technology, under a contract with the National Aeronautics and Space Administration.


\section*{References}


\end{document}